\documentclass[a4paper,11pt,english]{article}
\usepackage{makeidx,showidx}
\usepackage{latexsym}
\usepackage{amsmath}
\usepackage{amsfonts}
\usepackage{amssymb}
\usepackage{amsthm}
\usepackage{calc}
\usepackage[english]{babel}
\normalsize
\theoremstyle{definition}        %Bold header and normal body
\newtheorem{Def}{Definition}[section]

\theoremstyle{plain}              %Bold header and italic body
\newtheorem{Th}[Def]{Theorem}

\theoremstyle{remark}             %Italic header and normal body

\newtheorem*{ack}{Acknwoledgements}
\newcommand{\algeb}[1]{\mathfrak{#1}}

\newcommand{\R}{\mathbb{R}}

\newcommand{\N}{\mathbb{N}}
\newcommand{\la}{{\lambda}}                     %lambda
\newcommand{\al}{\alpha}
\newcommand{\eps}{\varepsilon}
\newcommand{\La}{\Lambda}

\DeclareMathOperator{\End}{End}
\DeclareMathOperator{\supp}{supp}

\DeclareMathOperator{\AD}{Ad}
\DeclareMathOperator{\Spp}{Sp}

\newcommand{\Id}{\mathbb{I}}
\newcommand{\rest}{\upharpoonright}
\newcommand{\cont}{\subseteq}
\newcommand{\trat}{\mspace{3mu}\negmedspace \text{--} \negthickspace \negthickspace}
\newcommand{\Oc}{{\mathcal{O}}}
\newcommand{\W}{{\mathcal{W}}}
\newcommand{\Hc}{{\mathcal{H}}}
\newcommand{\A}{\algeb{A}}
\newcommand{\AO}{\A(\Oc)}
\newcommand{\F}{\algeb{F}}                    %Alg. di campo glob. 
\newcommand{\FO}{\F(\Oc)}
\newcommand{\Pg}{\mathcal{P}}

\newcommand{\Pport}{\Pg_+^\uparrow}
\newcommand{\Lx}{ (\Lambda , x) }                %(Lambda,x)

\newcommand{\aLx}{{\alpha}_{\Lx}}                %alpha_{Lambda,x}

\newcommand{\ULx}{U(\Lambda , x)}                %U(Lambda,x)
\newcommand{\Vp}{V_+}
\newcommand{\Vpc}{\overline{\Vp}}
\newcommand{\Om}{\Omega}
\newcommand{\om}{\omega}                         %(mio)
\newcommand{\Aau}{\underline{{\A}}}             %global sc. algebra
\newcommand{\AOu}{\underline{{\A}}({\Oc})}
\newcommand{\Au}{\underline{A}}                  %element of sc. algebra
\newcommand{\Aul}{\underline{A}(\lambda)}        %value of "
\newcommand{\Ffu}{\underline{{\F}}}              %global sc. algebra
\newcommand{\FOu}{\underline{{\F}}({\Oc})}
\newcommand{\aLlx}{\alpha_{(\Lambda,\lambda x)}}
\newcommand{\alx}{{\alpha}_{\lambda x}}          %alpha_lambda x  (mio)
\newcommand{\au}{\underline{\alpha}}
\newcommand{\auLx}{\au_{\Lx}}
\newcommand{\oul}{{\underline{\omega}_\lambda}}  %omega sc._lambda   (mio)
\newcommand{\lak}{{\lambda_\kappa}}
\newcommand{\SL}{\operatorname{SL}}
\newcommand{\ou}{{\underline{\omega}}}           %omega sc.
\newcommand{\ouz}{\underline{\om}_0}
\newcommand{\oz}{\om_0}
\newcommand{\Oz}{\Om_0}

\newcommand{\Az}{\A_0}                 %algebra A_zero
\newcommand{\AzO}{\A_0(\Oc)} 
\newcommand{\Fz}{\F_0}

\newcommand{\pz}{\pi_0}                          %pi_zero
\newcommand{\Hz}{\mathcal{H}_0}                  %H_zero
\newcommand{\Uz}{U_0}
\newcommand{\UzLx}{\Uz(\Lambda , x)}                %U(Lambda,x)
\newcommand{\psjl}{\psi_j(\la)}

\newcommand{\auhpsj}{\au_h \psi_j}
\newcommand{\auhnpsj}{\au_{h_n} \psi_j}
\newcommand{\auhnpsl}{\au_{h_n} \psi_l}
\newcommand{\pszj}{\psi_{0,j}}

\newcommand{\Un}{\underline{U}_n}
\newcommand{\Um}{\underline{U}_m}
\newcommand{\Unl}{\Un(\la)}

\newcommand{\Uml}{\Um(\la)}

\newcommand{\ACTC}{\operatorname{ACTC}}
\newcommand{\ACTCt}{\ACTC_t}
\newcommand{\Ad}{\A^d}
\newcommand{\AdO}{\Ad(\Oc)}
\newcommand{\Azd}{\Az^d}
\newcommand{\AzdO}{\Azd(\Oc)}
\newcommand{\DHR}{\operatorname{DHR}}
\newcommand{\CTC}{\operatorname{CTC}}
\newcommand{\CTCt}{\CTC_t}
\newcounter{propcount}
\newlength{\proplength}
\newenvironment{proplist}[2][1]{\begin{list}{(\roman{propcount})}{\usecounter{propcount}\setcounter{propcount}{#2}\settowidth{\proplength}{\textit{(\roman{propcount})}}\setcounter{propcount}{#1-1}\setlength{\leftmargin}{\proplength + 0.5 em}\setlength{\rightmargin}{0 cm}\setlength{\topsep}{0 cm}\setlength{\itemsep}{0 cm}\setlength{\parsep}{\parskip}\setlength{\labelwidth}{\proplength}\setlength{\labelsep}{0.5 em}\setlength{\itemindent}{0 cm}}}{\end{list}}
\begin{document}
\begin{center}
{\Large\textbf{The Structure of Charges in the Ultraviolet\\
\vspace{0.3\baselineskip}and an Intrinsic Notion of Confinement}}\\
\vspace{1cm}
{\large Gerardo Morsella\\
\vspace{0.3cm}
\textit{Dipartimento di Matematica \\
Universit\`a di Roma ``La Sapienza''\\
P.le Aldo Moro, 2\\
I-00185 Roma, Italy}\\
\texttt{morsella@mat.uniroma1.it}}\\
\vspace{0.5cm}
27th May 2002\\
\end{center}
\vspace{0.5cm}
\begin{abstract}
After a brief account of the algebraic version of renormalization group developed by Buchholz and Verch, and of the main results of the Doplicher-Haag-Roberts theory of superselection sectors, we introduce the notion of asymptotic charge transfer chain, through which it is possible to reconstruct the scaling limit theory's superselection structure entirely in terms of underlying theory's observables. Furthermore, these objects allow the formulation of a natural notion of preservation of a charge in the scaling limit, so that one gets an intrinsic definition of confined charges, as those charges of the scaling limit which do not come from preserved charges of the underlying theory. 
\end{abstract}

\section{Introduction and statement of the problem}
A commonly accepted feature of quantum chromodynamics (QCD), the theory that is universally believed to describe hadronic physics phenomenlogy, is that it can be interpreted, at small spatio-temporal scales, in terms of particle-like structures, quarks and gluons, carrying a colour charge, which do not appear in the spectrum of physical states, due to the existence of a force between them that grows with distance. This is the phenomenon of \emph{confinement}. However, such a description may not be intrinsic, since it is based on the attachment of a physical interpretation to the unobservable Dirac and Yang-Mills fields out of which the theory is constructed, identified, respectively, with quark and gluon fields. It may well be the case that there exists another description of the theory, based on a completely different set of basic fields, which yields the same $S$ matrix and the same observables, yet not admitting an interpretation in terms of confined particles or charges. Several examples are known of such a situation. It is well known, for instance, that the algebra of observables of the Schwinger model (massless QED in 2 spacetime dimensions), is isomorphic to the algebra generated by a free massive scalar field \cite{Lowenstein:1971fc}, and therefore in the physical Hilbert space of the theory there appear no charged states, though a charged Dirac field enters in the Lagrangian. One would then be led to the interpretation according to which the theory describes a confined charged particle. However, one could have started simply with a free field Lagrangian obtaining the same final result, and this setting would leave no space for an interpretation in terms of confinement. Other examples have become popular in recent years, through the discovery of a web of dualities between supersymmetric Yang-Mills theories: in all such cases one has a couple of theories describing the same observables, but constructed in terms of completely different sets of fields and gauge groups. Based on the above considerations, D. Buchholz \cite{Buchholz:1996xk} has advocated the following point of view: in order to decide if the theory intrinsically describes entities which have to be identified with quarks, gluons and colour, and in order to have an intrisic notion of their confinement, one has to look at the observables alone.

The algebraic approach to quantum field theory \cite{Haag:1996a} is the most suitable one to address this problem. Indeed, on one hand, in this framework one has a completely general procedure which allows to reconstruct, from the knwoledge of the algebras of local observables, the complete set of charges of the theory (its \emph{superselection sectors}), as well as charge carrying fields and the global gauge group \cite{Doplicher:1971a, Doplicher:1974at, Doplicher:1990a}. On the other hand, through the intrinsic version of the renormalization group given by Buchholz and Verch \cite{Buchholz:1995a}, it is possible, still assuming only the knowledge of the local observables, to perform in a canonical way the scaling (ultraviolet) limit of a theory, which is again a theory formulated within the algebraic framework.

Using these tools, it is possible to give the following intrinsic notion of confinement of charges: a theory describes confined charges if there are charges of its scaling limit, which do not appear as charges of the theory itself.

However, this poses immediately the problem of developing a canonical way to compare the superselection structures (i.e. the set of charges) of the two theories, in such a way to be able to identify those charges of the scaling limit which are also charges of the underlying theory. In this contribution, we address this problem. In sect. \ref{sec:renormgroup} we shall recall the above mentioned algebraic version of renormalization group, based on the scaling algebra, and the construction of the (ultraviolet) scaling limit theory. In sect. \ref{sec:superstruc} we shall give a short account of the main results of the theory of superselection sectors, and of its formulation in terms of charge transfer chains. Finally, in sect. \ref{sec:confin} we shall give an ``asymptotic'' version of charge transfer chains, and we shall announce results which show that, using these objects, it is possible to reconstruct the superselection structure of the scaling limit theory in terms of observables of the underlying theory. Moreover, it will follow from our results that, using such asymptotic charge transfer chains, it is possible to identify in a natural way those charges of the scaling limit theory which are, in an appropriate sense, the scaling limit of charges of the underlying theory, thereby providing the above mentioned comparison betweeen the two superselection structures. The proof of these results will be published elsewhere. Some work in this same direction is also being done by C. D'Antoni and R. Verch \cite{D'Antoni:2001a}.

\begin{ack} I am grateful to my advisor, S. Doplicher, for having given me the opportunity to work on this subject and for his constant help, and to D. Buchholz and R. Verch for many useful discussions and suggestions, and for their warm hospitality, during some stage of this work, at the Institut f\"{u}r Theoretische Physik in G\"{o}ttingen, which I also acknowledge for financial support. 
\end{ack}

\section{Renormalization group and ultraviolet limit in Local Quantum Field Theory}
\label{sec:renormgroup}
In this section, we shall briefly recall the results of \cite{Buchholz:1995a} about the intrisic construction of the scaling limit of a quantum field theory in the framework of the theory of algebras of local observables \cite{Haag:1996a} -- also called \emph{Local Quantum Field Theory} (LQFT, for short). In this setting, a specific theory is defined by a correspondence
\begin{equation}
\label{eq:OinAO}
\Oc \to \AO
\end{equation}
between open bounded regions $\Oc \subset \R^4$ in Minkowski space-time, and (unital) C$^*$-algebras $\AO$, thought of as being generated by the observables of the theory which are measurable by an experiment performed in the region $\Oc$. By fixing a (pure) vacuum state, and going to its GNS representation, one can assume that all these C$^*$-algebras act on a \emph{vacuum Hilbert space} $\Hc$. Then, the correspondence (\ref{eq:OinAO}) is assumed to satisfy:
\begin{proplist}{4}
\item if $\Oc_1 \cont \Oc_2$, then $\A(\Oc_1) \cont \A(\Oc_2)$, i.e. $\Oc \to \AO$ is a net;
\item the net $\A$ is \emph{local}: if $\Oc_1$ is spacelike separated from $\Oc_2$, $\Oc_1 \cont \Oc_2'$ in symbols, then the algebras $\A(\Oc_1)$ and $\A(\Oc_2)$ commute, $\A(\Oc_1) \cont \A(\Oc_2)'$; 
\item the net $\A$ is \emph{Poincar\'e covariant}: there exists on $\Hc$ a unitary, strongly continuous representation $\Lx \to \ULx$ of the (proper orthocronous) Poincar\'e group $\Pport$, which induces a group $\aLx := \AD \ULx$ of automorphisms of the \emph{quasi-local algebra} $\A := \overline{\bigcup_\Oc \AO}$ (closure in the operator norm topology) such that
\begin{equation*}
\aLx(\AO) = \A(\La \Oc +x);
\end{equation*}
furthermore the translations satisfy the \emph{spectrum condition}
$\Spp U(\Id, \cdot) \cont \Vpc$ (the closed forward light cone), and there exists a unique (up to a phase) translation invariant vector $\Om$ (\emph{vacuum vector}), which is also cyclic for $\A$;
\item for every $A \in \A$ the function $\Lx \to \aLx(A)$ is norm
  continuous, and the local algebras are maximal with respect to this
  property, i.e. every $A \in \AO^-$ (weak closure) for which the
  above function is continuous, is already contained in $\AO$.
\item the net $\A$ satisfies \emph{geometric modular action}: let $(\Delta, J)$ be the pair associated by Tomita-Takesaki modular theory to $(\A(\W_+)^-, \Om)$, where $\W_+ := \{ x \in \R^4 : x^1 > |x^0| \}$ is the right wedge. Then
\begin{align*}
J \A(\Oc)^- J &= \A(j\Oc)^-, \\
J \ULx J &= U(j\La j, jx), \\
\Delta^{it} &= U(\La_{2 \pi t}),
\end{align*}
where $(\La_s)_{s \in \R}$ is the one parameter group of boosts in the $x^1$ direction, and $j$ is the reflection in the $(x^0,x^1)$ plane.
\end{proplist}

For a discussion of the physical motivations of the assumptions (i)-(iii),
as well as for a comprehensive overview of the understanding of
structural properties of quantum field theory gained through them, we
refer the reader to \cite{Haag:1996a}. For what concerns assumption
(iv), which is crucial for the construction of the scaling limit, we
only remark that it is not really restrictive, as discussed in
\cite{Buchholz:1995a}. Finally, assumption (v) is verified, e.g., if the local algebras can be derived from underlying Wightman fields \cite{Bisognano:1976za}, and it can be proved at a purely algebraic level under general assumptions \cite{Borchers:1992xk, Brunetti:1993zf, Mund:2001sv}. 

As remarked in \cite{Buchholz:1995a} a common feature of all possible
families of renormalization group (RG) transformations $(R_\la)_{\la > 0}$, in the usual
(lagrangean) approach to quantum field theory, is that they map observables localized in $\Oc$ to observables localized in $\la \Oc$ (since the speed of light $c$ has to remain constant), and observables which transfer to states 4-momentum contained in a region $\tilde{\Oc}$ to observables which transfers 4-momentum in $\la^{-1} \tilde{\Oc}$ (since also Planck's $\hbar$ has to remain constant). Equivalently, the RG orbits $\la \to R_\la(A)$ have the following continuity property with respect to Poincar\'e transformations \cite[lemma 3.1]{Buchholz:1995a}:
\begin{equation}
\lim_{\Lx \to (\Id, 0)} \sup_{\la > 0} \| \aLlx(R_\la(A)) - R_\la(A) \| = 0.
\end{equation}

As discussed at length in \cite{Buchholz:1995a}, according to the
principles of LQFT, what really matters are only the above stated
phase space properties of RG orbits, and we are thus led to the
following

\begin{Def}
On the C$^*$-algebra of bounded functions $\la \to \Aul \in \A$, with the norm
\begin{equation*}
\| \Au \| := \sup_{\la > 0} \| \Aul \|,
\end{equation*}
and pointwise defined algebraic operations, we get an action $\au$ of $\Pport$ as
\begin{equation}
\auLx(\Au)(\la) := \aLlx(\Aul).
\end{equation}
Then the \emph{local scaling algebra} relative to region $\Oc$ is the C$^*$-algebra $\AOu$ of all bounded functions $\Au$ such that $\Aul \in \A(\la\Oc)$ and
\begin{equation}
\lim_{\Lx \to (\Id,0)} \| \auLx(\Au) - \Au \| = 0.
\end{equation}
The \emph{(quasi-local) scaling algebra} $\Aau$ is the C$^*$-inductive limit of the net $\Oc \to \AOu$.
\end{Def}

It is then clear that $\Aau$ is a Poincar\'e covariant, local net of C$^*$-algebras. With this tool at hand, we can study the properties of physical states of the \emph{underlying theory} $\A$ in the limit of short distances (i.e. high energies). For that, given a locally normal state $\om$ on $\A$ we can define its lift to $\Aau$ as the family of states
\begin{equation}
\oul(\Au) := \om(\Aul), \qquad \la > 0, \Au \in \Aau.
\end{equation}
We will regard $(\oul)_{\la > 0}$ as a  net directed by $\la \to 0$, and we shall denote by $\SL(\om)$ the set of its weak*-limit points, which is non-void by Banach-Bourbaki-Alaoglu theorem. As a consequence of the fact that, for any two locally normal states, $\| (\om_1-\om_2)\rest\A(\la\Oc) \| \to 0$ for $\la \to 0$ \cite{Roberts:1974aa}, and of clustering estimates in \cite{Araki:1962aa} one has

\begin{Th}\emph{\cite[sect. 4]{Buchholz:1995a}} $\SL (\om)$ is
  independent of $\om$. Let $\ouz \in \SL (\om)$ with GNS
  representation $(\pz, \Hz, \Oz)$, and define $\AzO := \pz(\AOu)$,
  $\oz := (\Oz | (\cdot) \Oz)$. Then $\Oc \to \AzO$ is a local net of
  C$^*$-algebras, covariant with respect to a suitable representation
  $\Lx \to \UzLx$ of $\Pport$ satisfying the spectrum condition and
  leaving $\Oz$ invariant. $\oz$ is pure\footnote{All these statement
    but the last one, are true in any number of spacetime dimensions
    $d$. The scaling limit vacuum $\oz$ is pure only for $d \geq 3$}.
\end{Th}

Every net $\Az$ arising as in the above theorem, will be called a
\emph{scaling limit net} of $\A$. We see that there is the possibility
of a non-uniqueness of the scaling limit theory, since it may happen
that the theory varies continually as $\la$ approaches 0. We can
expect this to be the case for theories that, in the conventional
setting, do not admit an ultraviolet fixed point. On the other hand,
we can expect that the scaling limits of theories having an ultraviolet fixed point (in particular asymptotically free ones) will all be isomorprhic (as nets). This is the case, for instance, for the theory of a free massive scalar field in $d \geq 3$, for which the scaling limits are all isomorphic to the net generated by the free massless scalar field \cite{Buchholz:1998vu}, and also  for dilatation invariant theories \cite[sect. 5]{Buchholz:1995a} (which satisfy the Haag-Swieca compactness condition \cite{Haag:1965aa}).

The above construction of the scaling limit relies only on assumptions
(i)-(iv). Assumption (v) is however fundamental in analysing the
scaling limit superselection structure: if $\A$ satisfies (v), so does any scaling limit net $\Az$ \cite[sect. 6]{Buchholz:1995a}, which implies \emph{essential Haag duality} in the scaling limit
\begin{equation}
\Azd(\Oc')' = \AzdO,
\end{equation}
where $\AzdO := \Az(\Oc')'$ is the \emph{dual net} of $\Az$. This is a fundamental issue for the DHR theory of superselection sectors, which we are going to describe.

\section{Superselection structure and charge transfer chains}
\label{sec:superstruc}
Superselection sectors have been one of the first and most succesful applications of LQFT. Here we can give only a brief account of the results which are useful for us, referring to the original papers for details. According to Haag and Kastler \cite{Haag:1964dh}, given a local net $\Oc \to \AO$, the superselection sectors of the theory have to be identified with unitary equivalence classes of (irreducible) representations of the quasi-local algebra. However, one expects that only a subclass of representations describes states which are relevant for particle physics. Restricting then to states which are, in some sense, local excitations of the vacuum, one arrives at the following selection criterion for representations of $\A$ \cite{Doplicher:1971a}.

\begin{Def}
The representation $\pi$ of the quasi-local algebra $\A$ satisfies the \emph{DHR selection criterion} if, for every double cone $\Oc \cont \R^4$
\begin{equation*}
\pi \rest \A(\Oc') \cong \pz \rest \A(\Oc')
\end{equation*}
(unitarily equivalent), $\pz$ begin the defining (vacuum) representation of $\A$.
\end{Def}

The corresponding charges (i.e. labels of unitary equivalence classes)
are termed \emph{localizable charges}. This excludes theories in which
long range forces are present, such as QED: due to Gauss' law, the
electric charge is measurable in the spacelike complement of any
bounded region. Moreover, also in purely massive theories there are
\emph{topological charges} for which the above criterion only holds if
one replaces double cones by spacelike cones
\cite{Buchholz:1982fj}. For such charges it is possible to develope a
superselection theory which is completely analogous to the one for
localizable charges. However localizable charges are the only ones
expected to occur in the scaling limit in physically interesting cases
(asymptotically free theories).

The representations complying with the above criterion form the objects of a C$^*$-category, whose arrows are the intertwiners between representations. If we assume that the net $\A$ satisfies essential Haag duality and a consequence of weak additivity\footnote{A net $\A$ is said to be \emph{weakly additive} if, for every double cone $\Oc$, the von Neumann algebra generated by $\A(\Oc + x)^-$, $x \in \R^4$, is the algebra of all bounded operators on the vacuum Hilbert space.} and positivity of the energy, known as property B (for which we refer to \cite{Doplicher:1971a}), then this category is equivalent to the C$^*$-category $\DHR(\Ad)$ of all localized transportable endomorphisms of $\Ad$ and their intertwiners \cite{Doplicher:1971a, Roberts:1989ps}, where a $\rho \in \End(\Ad)$ is called \emph{localized} in $\Oc$ if $\rho(A) = A$ for every $A \in \Ad(\Oc')$, and it is called \emph{transportable} if for every $\tilde{\Oc}$ there is a $\tilde{\rho}$ localized there, which is equivalent to $\rho$. As shown by Doplicher and Roberts \cite{Doplicher:1990a}, this category has the much richer structure of a symmetric tensor C$^*$-category with subobjects, direct sums and conjugates, which allows the reconstruction of a field net $\Oc \to \FO$, satisfying normal Bose-Fermi commutation relations, and, acting on it, a compact gauge group $G$, in such a way that $\Ad$ is the fixed point subnet of $\F$ under the action of $G$.

There is also an alternative description, essentially due to Roberts
\cite{Roberts:1989ps}, of superselection structure in terms of
so-called \emph{charge transfer chains}, which we are going to discuss (without entering into details). We say that an observable $A$ is \emph{bilocalized} in a couple of double cones $\Oc_1$ and $\Oc_2$ if $A \in \AdO'$ for every double cone $\Oc \cont \Oc_1' \cap \Oc_2'$.

\begin{Def}
A \emph{charge transfer chain} is a sequence $(U_n)_{n \in \N}$ of unitaries in $\Ad$, such that there exists a sequence of double cones $(\Oc_n)_{n \in \N_0}$, eventually spacelike to any given double cone, in such a way that $U_n$ is bilocalized in $\Oc_0$ and $\Oc_n$, and $U_m^* U_n^{\phantom{*}}$ is bilocalized in $\Oc_m$ and $\Oc_n$.
\end{Def}

We may think of $U_m^* U_n^{\phantom{*}}$ as the operation of shifting a charge from $\Oc_n$ to $\Oc_m$, as we are going to see. Transporable charge transfer chains (obviously definded) are the objects of a C$^*$-category $\CTCt(\Ad)$, whose arrows are suitable sequences $(T_n)_{n \in \N} \cont \Ad$.

\begin{Th} \label{thm:ctc}
There is an equivalence of C$^*$-categories $\Phi : \CTCt(\Ad) \to \DHR(\Ad)$ given, on the objects, by $(U_n)_{n \in \N} \to \rho^U$, with
\begin{equation*}
\rho^U(A) = \lim_{n \to +\infty} U_n^{\phantom{*}} A U_n^*,
\end{equation*}
the limit existing in norm. $\rho^U$ is localized in $\Oc_0$.
\end{Th}

This implies, in particular, that there is a bijection between superselection sectors and equivalence classes of charge transfer chains.

\section{Superselection structure in the ultraviolet and confinement}
\label{sec:confin}
In view of the results recalled in the above sections, the structure of charges of the scaling limit theory $\Az$ has to be regarded as an intrisic feature of the underlying theory $\A$. Thus, if one had a canonical way to identify charges of the underlying theory with (a subset of) charges of the scaling limit theory, one would get an intrisc notion of confinement: confined charges are those charges of the scaling limit theory which do not come from charges of the underlying theory. A simple example of this situation is provided by the already mentioned Schwinger model: as recalled in the introduction, the algebra of observables of this model is isomoprhic to the algebra generated by a single free massive scalar field \cite{Lowenstein:1971fc}, and therefore it has no superselection sectors (apart from the vacuum). However, the algebra $\Az$ in the scaling limit is (a local extension of) the algebra generated by the free massless scalar field (in Weyl form), and it has a one-parameter family of (cone-like localizable) superselection sectors, carrying an ``electric'' charge \cite{Buchholz:1996xk, Buchholz:1998vu}, which is therefore confined. Since in this model the underlying theory describes no charges, it is natural to call the scaling limit charges the confined ones. In more complicated situations, however, it is apparent that a way to compare the superselection structure of the two theories is needed. To make such a comparison, what one would need is a natural notion of preservation of charge in the scaling limit, to which we turn now.

From the results of the above sections, we see that the superselection structure of the scaling limit can be described by sequences $(U_n)_{n \in \N} \cont \Azd$ performing the $n \to +\infty$ limit, and elements of $\Az$ are obtained (morally) as limits, for $\la \to 0^+$, of functions $\la \to \Aul \in \A$. It should then be possible to describe the superselection structure of $\Az$ by suitable families of observables $\Unl$, $n \in \N$, $\la > 0$, and performing the double limit $\la \to 0^+$, $n \to +\infty$. This is indeed possible, as we are going to see.

Let $\A$ be a net satisfying (i) - (v), if we also assume that $\A$ complies with a suitable version of the nuclearity condition \cite{Buchholz:1986dy}, it follows \cite{Buchholz:1996mx} that each scaling limit theory is also (suitably) nuclear, and, in turn, that it has property B. Let then $\Az$ be one of such scaling limits, arising as the GNS representation of $\ouz \in \SL(\om)$, and let $(\ou_{\la_\kappa})_{\kappa \in I}$ be a subnet which has $\ouz$ as its weak* limit. 

\begin{Def}
An \emph{asymptotic charge transfer chain} is a bounded sequence $(\Un)_{n \in \N} \cont \Aau(\W)$, where $\W$ is some translate of the right wedge, such that
\begin{proplist}{3}
\item there holds
\begin{equation*}
\lim_{n \to +\infty} \lim_{\kappa \in I} \| [\Un(\lak)^*\Un(\lak)-\Id]\Om \| = 0 = \lim_{n \to +\infty} \lim_{\kappa \in I} \| [\Un(\lak)\Un(\lak)^*-\Id]\Om \|;
\end{equation*}
\item there exists a double cone $\Oc_0 \cont \W$ such that for every $\Au \in \Aau(\Oc_0')$
\begin{equation*}
\lim_{n \to +\infty} \lim_{\kappa \in I} \| [\Un(\lak) \Au(\lak) - \Au(\lak) \Un(\lak)]\Un(\lak)^* \Om \| = 0;
\end{equation*}
\item for all double cones $\Oc$ and all $\A \in \AOu$
\begin{equation*}
\lim_{m,n \to +\infty} \lim_{\kappa \in I} \| [\Um(\lak)^* \Un(\lak) \Au(\lak) - \Au(\lak) \Um(\lak)^* \Un(\lak) ] \Un(\lak)^* \Om \| = 0.
\end{equation*}
\end{proplist}
\end{Def}

What the two last conditions above express is essentially that $\Unl$ is, asymptotically as $\la \to 0^+$ and $n \to +\infty$, bilocalized in a couple of regions, one of which is $\Oc_0$, while the other one goes to spacelike infinity, and that $\Uml^* \Unl$ is asymptotically bilocalized in a couple of regions both going to spacelike infinity (and in the same direction). This is analogous, though weaker, to the localization properties of usual charge transfer chains.

In analogy with theorem \ref{thm:ctc}, through asymptotic charge transfer chains it is possible to reconstruct the scaling limit's superselection structure.

\begin{Th}
Transportable asymptotic charge transfer chains are the objects of a C$^*$-category $\ACTCt(\A)$ (whose arrows are suitable sequences $(\underline{T}_n)_{n \in \N} \cont \Aau$). $\ACTCt(\A)$ is equivalent to $\DHR(\Azd)$, and the equivalence is given, on the objects, by $(\Un)_{n \in \N} \to \rho^{\underline{U}}$, with
\begin{equation*}
\rho^{\underline{U}}(A) = s\mspace{0.5mu}\trat\mspace{-1.0mu}\lim_{n \to +\infty} \pz(\Un)A\pz(\Un)^*, \qquad A \in \Az
\end{equation*}
(limit in the strong operator topology). $\rho^{\underline{U}}$ is localized in $\Oc_0$.
\end{Th}

Using asymptotic charge transfer chains it is possible to compare the superselection stuctures of $\A$ and $\Az$, essentially by looking at those charges for which it is possible to interchange the two limits, $\la \to 0^+$ and $n \to +\infty$. By this we mean that it may well happen that for some scaling limit charge $\xi_0$, the corresponding asymptotic charge transfer chain could, in some sense, also create a fixed charge $\xi$ at each finite scale $\la > 0$. It would then be natural to identify the charge $\xi$ of $\A$ and $\xi_0$ of $\Az$, i.e. to regard $\xi_0$ as the scaling limit of the charge $\xi$. Then, $\xi_0$ would be a non-confined charge.  In order to make precise the above idea of scaling limit of charge, we turn to the consideration of the scaling limit of charge carrying fields.

Suppose that the superselection structure of $\A$ is described by a
translation covariant\footnote{For simplicity, we do not assume full
  Poincar\'e covariance of the field net, as this would result in
  unnecessary complications of the following definitions of smeared
  charged multiplets and of UV-stable charges.} field net $\Oc \to
\FO$ satisfying normal Bose-Fermi commutation rules, and by a compact
gauge group $G$, acting locally on $\F$ and such that $\AO^-$ is the
fixed point subalgebra of $\FO$ under this action, as in
\cite{Doplicher:1990a}. In this situation an endomorphism $\rho$ of
$\A$ localized in $\Oc$ is implemented by a multiplet of orthogonal
isometries of support $\Id$ in the field net, i.e. there exist field operators $\psi_j \in \FO$, $j = 1, \dots, d$, such that
\begin{equation}
\psi_l^* \psi_j^{\phantom{*}} = \delta_{l,j}\Id, \qquad \sum_{j=1}^d \psi_j^{\phantom{*}} \psi_j^* = \Id, \qquad \rho(A) = \sum_{j=1}^d \psi_j^{\phantom{*}} A \psi_j^*,
\end{equation}
and the $\psi_j$'s transform according to an irreducible representation of $G$.

It is easy to see that it is possible to define the field scaling algebra $\Ffu$ and the field algebra scaling limit $\Fz$ in complete analogy to what we have done for the observable algebra.

Let $\xi$ be a charge of $\A$, and, for each $\la > 0$, let $\rho_\la$ be an endomorphism of class $\xi$ of $\A$ localized in $\la \Oc$, and $\psjl \in \F(\la \Oc)$, $j=1,\dots,d$, the corresponding charge multiplet. In general the function $\la \to \psjl$ will not be an element of $\FOu$, as it lacks the right continuity properties with respect to the action of the translations. Following \cite{D'Antoni:2001a} we define \emph{smeared charge multiplets}
\begin{equation}
(\auhpsj)_\la := \int_{\R^4} d^4x \, h(x) \alx(\psjl),
\end{equation}
with $h \in L^1(\R^4)$, $\int h = 1$, and it is then easy to see that $\auhpsj \in \Ffu(\Oc + \supp h)$. We now want to look at the scaling limit of these charge multiplets: if these scaling limits are charge multiplets themselves, they create a charge of $\Az$, which is natural to identify as the scaling limit of the charge $\xi$. However, due to the smearing with the function $h$, in general one cannot hope that the operators $\pz(\auhpsj)$ form a multiplet of orthogonal isometries. Worse than that, if $(h_n)_{n \in \N} \cont L^1(\R^4)$ is a $\delta$-sequence (e.g. $h_n \geq 0$ has support in a ball of radius $1/n$ centered in $0$, and $\int h_n = 1$), then there exists the limit
\begin{equation}
\label{eq:limmult}
s^* \trat \lim_{n \to +\infty} \pz(\auhnpsj) =: \pszj \in \Fz(\Oc_0)^-
\end{equation}
where $\Oc_0$ is any double cone containing the closure of $\Oc$, but still, in general, $(\pszj)_{j=1,\dots,d}$ is not a charge multiplet (e.g. it may be zero). This is to be expected: charges may disappear in the scaling limit, for instance due to an exceptional quantum behaviour of the associated fields, as may be the case in theories without an ultraviolet fixed point (cfr. the discussion in \cite{Buchholz:1996xk, Buchholz:1995a} of the case of ``classical'' scaling limit). What one needs, in order to get a nontrivial scaling limit of charge multiplets, is a condition expressing the physical fact that localizing the charge in smaller and smaller regions does not require too much energy.

\begin{Def}
The charge $\xi$ is \emph{UV-stable} if there is a family of associated charge multiplets $(\psjl)_{j=1,\dots,d} \cont \F(\la\Oc)$, $\la > 0$, such that for every $\eps > 0$ there is a compact $\Delta \cont \R^4$ for which
\begin{equation*}
\sup_{\la > 0} \| [E(\la^{-1}\Delta)-\Id]\psjl\Om\| < \eps,
\end{equation*}
where $E$ is the spectral measure determined by the translations.
\end{Def}

What we are requiring is then essentially that for an UV-stable charge, energy of order $\la^{-1}$ is needed to be able to create the charge from the vacuum in a region of radius of order $\la$.\footnote{As angular momentum has the dimensions of $\hbar$, and therefore it is not running under RG, in the case of a Poincar\'e covariant theory, in order to get a nonvanishing charge in the scaling limit, it would be necessary to add a condition expressing the fact that $\psjl \Om$ has angular momentum independent of $\la$, uniformly as $\la \to 0$.} This condition of preservation of charge is analogous, though apparently somewhat stronger, to that formulated, for the same purpose, by D'Antoni and Verch \cite{D'Antoni:2001a}. We also remark that the above condition is verified in free field models.

\begin{Th}
With the notations introduced above, let $\xi$ be an UV-stable charge of $\A$. Then $(\pszj)_{j=1,\dots,d}$, defined as in (\ref{eq:limmult}), is a multiplet of orthogonal isometries of support $\Id$. Moreover, if $(x_n)_{n \in \N} \cont \W$ is a sequence going to spacelike infinity and if we define
\begin{equation*}
\Unl := \sum_{j=1}^d (\auhnpsj)_\la \al_{\la x_n}((\auhnpsl^*)_\la), \qquad n \in \N, \la > 0,
\end{equation*}
then $(\Un)_{n \in \N}$ is an asymptotic charge transfer chain, such that
\begin{equation}
\label{eq:rho_la}
\rho_\la(A) = s\mspace{0.5mu}\trat\mspace{-1.0mu}\lim_{n \to +\infty} \Unl A \Unl^*, \qquad A \in \A, \la > 0,
\end{equation}
and
\begin{equation*}
\rho^{\underline{U}} (A)= s\mspace{0.5mu}\trat\mspace{-1.0mu}\lim_{n \to +\infty} \pz(\Un) A \pz(\Un)^* = \sum_{j=1}^d \pszj A \pszj^*, \qquad A \in \Az.
\end{equation*}
\end{Th}

Then, an UV-stable charge is preserved in the scaling limit, and there is a corresponding asymptotic charge transfer chain that ``creates'' the same charge at every finite scale. This motivates the following intrinsic notion of (non-)confinement, at least for theories without quantum topological charges.

\begin{Def}
A scaling limit charge $\xi_0$ is \emph{non-confined} if there exists an asymptotic charge transfer chain $(\Un)_{n\in\N}$ such that $\xi_0 = [ \rho^{\underline{U}}]$, and a family $(\rho_\la)_{\la > 0}$ of equivalent localized endomorphisms of $\A$, $\rho_\la$ localized in $\la \Oc$, such that (\ref{eq:rho_la}) holds.
\end{Def}

Phrased differently, a non-confined charge is a scaling limit charge which can also be created at each finite scale. In the general case of a theory describing also quantum topological charges \cite{Buchholz:1982fj}, which are localizable in arbitrary spacelike cones, one expects, on physical grounds, that non-confined cone-like localizable charges give rise to double cone localizable charges in the scaling limit, because of the fact that the infinite string attached to such a charge becomes weaker and weaker at small scales, and eventually disappears. So, we see that the above notion of confinement does not apply to these theories, as a double cone localizable charge $\xi_0$ being the scaling limit of a cone-like localizable charge $\xi$, would be confined according to it. Work is in progress in order to treat this more general case.

\end{document}